\documentclass{svjour3}  
\RequirePackage{fix-cm}
\usepackage[a4paper]{geometry}
\usepackage{amsmath,amssymb}
\usepackage{tikz,tkz-tab}
\usepackage{graphics}

\def\NN{{\mathbb{N}}}

\def\RR{{\mathbb{R}}}

\def\CC{{\mathcal{C}}}

\def\PP{{\mathcal{P}}}

 \newtheorem{prop}{Proposition}

\title{k-medoids and p-median clustering are solvable in polynomial time for a 2d Pareto front}
\author{Nicolas Dupin, Frank Nielsen, El-Ghazali Talbi\\
{dupin@lri.fr}
}

\date{
}

\begin{document}

 \title{k-medoids and p-median clustering are solvable in polynomial time for a 2d Pareto front}
 \titlerunning{p-median is  polynomial for a 2d Pareto front}
 \author{Nicolas Dupin
 \and Frank Nielsen
 \and El-Ghazali Talbi
 }
%
%
%
%
\institute{N. Dupin \at  Laboratoire de Recherche en informatique (LRI), CNRS, Universit\'e Paris-Saclay, France\\
\email{nicolas.dupin@universite-paris-saclay.fr}
\and
F. Nielsen \at Sony Computer Science Laboratories Inc, Tokyo, Japan\\  
\email{Frank.Nielsen@acm.org}\\
\and
E. Talbi \at
Univ. Lille, UMR 9189 - CRIStAL - Centre de Recherche en Informatique Signal et Automatique de Lille, F-59000 Lille, France\\
 \email{el-ghazali.talbi@univ-lille.fr}
}
\journalname{Preprint}

\date{
}

\maketitle

\begin{abstract}
This paper examines a common extension of k-medoids and k-median clustering in the  case of a two-dimensional Pareto front, as generated by bi-objective optimization approaches. 
A characterization of optimal clusters is provided, 
which allows to solve the optimization problems to optimality in polynomial time using a common dynamic  programming algorithm.
More precisely, having $N$ points to cluster in $K$ subsets, the complexity of the algorithm is proven  in $O(N^3)$ time and $O(K.N)$ memory space
when $K\geqslant 3$, cases $K=2$ having a time complexity in $O(N^2)$.
Furthermore, speeding-up the dynamic programming algorithm is possible  avoiding useless computations,
for a practical speed-up without improving  the complexity.
Parallelization issues are also discussed, to speed-up the algorithm in practice.


\vskip 0.2cm
\end{abstract}

\section{Introduction}

This paper is motivated by real-life applications of Multi-Objective Optimization (MOO).
Some optimization problems can be driven by more than one objective function, with some conflicts among objectives. 
For example, one may minimize financial costs, while maximizing the robustness to uncertainties \cite{dupin2015modelisation,peugeot2017mbse}.
In such cases, higher levels of robustness are likely to induce  financial over-costs.
Pareto dominance, preferring a solution from another if it is better for all the objectives, is a weak dominance rule.
With conflicting objectives, several non-dominated  solutions can be generated, these \emph{efficient} solutions are the best compromises.
A \emph{Pareto front} (PF) is the projection in the objective space  of the non-dominated solutions \cite{ehrgott2003multiobjective}.

MOO approaches may  generate large PF, for a trade-off evaluation by a decision maker. 
The problem is here to select  $K$  good compromise solutions  from $N\gg K$ non dominated solutions
while maximizing  the representativity of these $K$ solutions.
This problem can be seen as an application of clustering algorithms,
partitioning the $N$ elements into $K$ subsets with a minimal diversity,
and giving a representative element of the optimal clusters.
Selecting best compromise solutions for human  decision makers, one deals with small values $K<10$.
We note that partial PF are used inside population meta-heuristics \cite{talbi2009metaheuristics}.
Clustering a PF is also useful in this context to archive representative solutions of the partial PF \cite{zio2011clustering}. 
For such applications, the values of $K$ are larger than the previous ones. 

k-means clustering  is one of  the  most  famous   unsupervised  learning   problem,
and is widely studied in the literature since the seminal algorithm provided by Lloyd in \cite{lloyd1982least}.
The k-medoids problem, the discrete variant of the k-means problem, fits better with 
our application to maximize the dissimilarity around a representative solution \cite{kaufman1987clustering}.
If k-medoids clustering is more combinatorial than k-means clustering,
it is known to be more robust on noises and outliers  \cite{jain2010data}.
Both k-medoids and k-means problem  are NP hard in the general and the planar case \cite{aloise2009np,hsu1979easy}.
Lloyd's algorithm can be extended to solve heuristically k-medoids problems.
PAM (Partitioning Around Medoids), CLARA (Clustering LARge Applications) and CLARANS (Clustering Large Applications
based upon RANdomized Search) are such heuristics for k-medoids clustering \cite{schubert2018faster}.
Hybrid and genetic algorithms were also investigated in \cite{sheng2006genetic}.
Previous heuristics converge only to local minima, without any guarantee to reach a global minimum.
 This paper proves that the special case of k-medoids clustering in a two dimensional (2-d) PF
 is solvable in polynomial time, thanks to a Dynamic Programming (DP) algorithm.
Actually, the result is obtained for a common extension of k-medoids and k-median problems,
with common issues to speed up the DP   algorithm in practice.
Having $N$ points to cluster in $K$ subsets, the complexity of the algorithm is proven  in $O(N^3)$ time and $O(KN)$ memory space
when $K\geqslant 3$, cases $K=2$ having a time complexity in $O(N^2)$.
We note that the preliminary work \cite{dupin2019medoids} stated a complexity in $O(N^3)$ time and $O(N^2)$ memory space for the k-medoid problem.
 

In section 2, we define formally the problem and unify the notation.
In section 3, intermediate results and a characterization of optimal clusters are presented.
In section 4, it is described how to compute efficiently the costs of the previous clusters. 
 In section 5, a first DP algorithm is presented  with a proven polynomial complexity.
 In section 6,  it is discussed how to speed up the DP algorithm in practice, without improving the previous complexity. 
  In section 7, numerical and computational experiments are presented.  
In section 8,    our contributions  are summarized,  discussing also future directions of research.

  \begin{figure}[ht]
      \centering 
   \begin{tikzpicture}[scale=4]
    \draw[->] (-1,0) -- (1.1,0) node[right] {$x$};
    \draw[->] (0,-0.55) -- (0,0.7) node[above] {$y$};
    \draw (0.3,0) node[above] {$O=(x_O,y_O)$};
    \draw (0,0) node[color=black
    ] {$\bullet$};
    \draw (0.5,0.5) node[above] {$B=(x_B,y_B)$ with};
    \draw (0.5,0.4) node[above] {$x_O<x_B$ and $y_O<y_B$ };
    \draw (0.5,0.3) node[above] {$O$ dominates $B$};
    
    \draw (-0.5,-0.3) node[above] {$C=(x_C,y_C)$ with};
    \draw (-0.5,-0.4) node[above] {$x_O>x_C$ and $y_O>y_C$ };
    \draw (-0.5,-0.5) node[above] {$C$ dominates $O$};
    
    \draw (0.5,-0.3) node[above] {$D=(x_D,y_D)$ with};
    \draw (0.5,- 0.4) node[above] {$x_O<x_D$ and $y_D<y_O$ };
    \draw (0.5,- 0.5) node[above] {$O$ and $D$ not comparable};
    
    \draw (-0.5,0.5) node[above] {$A=(x_A,y_A)$ with};
    \draw (-0.5,0.4) node[above] {$x_O>x_A$ and $y_O<y_A$ };
    \draw (-0.5,0.3) node[above] {$A$ and $O$ not comparable};

\end{tikzpicture}
    \caption{Illustration of Pareto dominance and incomparability quarters minimizing two objectives indexed by $x$ and $y$: zones of $A$ and $D$ are incomparability zones related to $O$
    }\label{defIllustr}  
   \end{figure}
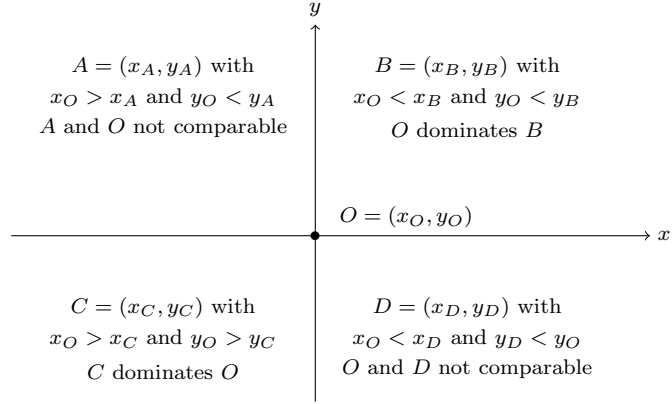

\section{Problem statement and notation}



We consider a set $E=\{x_1,\dots, x_N\}$ of $N$ elements of $\RR^2$, 
such that for all $ i\neq j$, $x_i \phantom{0} \mathcal{I} \phantom{0} x_j$  
defining the binary relations $\mathcal{I},\prec $  for all $ y=(y^1,y^2),z=(z^1,z^2) \in \RR^2$ with:
\begin{eqnarray}
 y \prec z  & \Longleftrightarrow  & y^1< z^1 \phantom{2} \mbox{and}\phantom{2} y^2> z^2\\
y \preccurlyeq z  & \Longleftrightarrow  &  y \prec z  \phantom{2} \mbox{or}\phantom{2} y= z\\
y \phantom{1}\mathcal{I}\phantom{1} z  & \Longleftrightarrow  &y \prec z  \phantom{2} \mbox{or} \phantom{2}  z \prec y 
\end{eqnarray}


%
%
%
%

These hypotheses on $E$  characterizes 2-d discrete PF minimizing two objectives,  as illustrated in Figure \ref{defIllustr}. We note that the convention leading to the definitions of $\mathcal{I},\prec$ considered the   minimization of two objectives.
 This is not a loss of generality, any bi-objective optimization problem can be transformed into a minimization of two objectives. 
Such a set $E$ can be extracted from any subset of $\RR^2$ using an output-sensitive algorithm  \cite{nielsen1996output},
or generated by  bi-objective optimization approaches, exact methods \cite{ehrgott2003multiobjective}
and also  meta-heuristics \cite{talbi2009metaheuristics}.

We consider in this paper the Euclidian distance : 
\begin{equation}\label{distEucl}
d(y,z) = |\!| y -z |\!| = \sqrt{ \left(y^1 - z^1\right)^2 + \left(y^2 - z^2\right)^2}, \hskip 0.1cm \forall  y=(y^1,y^2),(z^1,z^2) \in \RR^2
\end{equation}



\noindent{ Let $K\in\NN^*$ a strictly positive integer. $\Pi_K(E)$ denotes the set of the possible partitions of $E$ in $K$ subsets:

\begin{equation}
\Pi_K(E) = \left\{P \subset \PP(E)\: \bigg| \:\forall p,p' \in P, \:\:p \cap p' =  \emptyset \:, 
\: \bigcup_{p \in P} p = E \: \mbox{and} \; |P|=K \: \right\} 
\end{equation}

Defining a  cost function $f$ for each subset of $E$ to measure the dissimilarity, we investigate  $K$-clustering
problems which can be  written as following  combinatorial optimization problems,
minimizing the sum  of the  measure $f$ for all the $K$ clusters partitioning $E$: 
\begin{equation}\label{clusteringGal}
\min_{\pi \in \Pi_K(E)}  
\sum_{P \in \pi}  f(P)
\end{equation}

\emph{K-medoids} and \emph{K-median} problems are in the shape of (\ref{clusteringGal}).
On one hand, discrete \emph{K-median} cost function $f_{med}(P)$ considers the minimal   sum of the  distances from one chosen point  of $P$, denoted as the \emph{median}, to the other points of $P$.
On the other hand,
\emph{K-medoids} cost function $f_{mdd}(P)$ considers the minimal sum of the squared distances from one chosen point  of $P$, denoted as the \emph{medoid}, to the other points of $P$:

%
\begin{equation}\label{defKmedian}
\forall P  \subset E, \;\;\; f_{med}(P) = \min_{y \in P} \sum_{x \in P}  \left|\!\left| x - y \right|\!\right|
\end{equation}
\begin{equation}\label{defKmedoid}
\forall P  \subset E, \;\;\;
f_{mdd}(P)=\min_{c \in P}\sum_{x \in P}  \left|\!\left| x - c \right|\!\right|^2
\end{equation}

We unify notations  with $\alpha >0$, considering the generic dissimilarity function $f_{\alpha}(P)$:
\begin{equation}\label{defGeneric}
\forall P  \subset E, \;\;\; f_{\alpha}(P) = \min_{y \in P} \sum_{x \in P}  \left|\!\left| x - y \right|\!\right|^{\alpha}
\end{equation}

\begin{definition}[$\alpha$-medoids] For all $\alpha >0$,
the $\alpha$-medoid of a subset $P \subset E$ denotes the point $p \in P$ 
such that $p= \mbox{argmin}_{y \in P} \sum_{x \in P}  \left|\!\left| x - y \right|\!\right|^{\alpha}$
\end{definition}
\begin{definition}[$K$-$\alpha$-Med2dPF ]
For all $\alpha >0$,
the $K$-clustering problem  (\ref{clusteringGal}) using generic function $f_{\alpha}(P)$ is denoted $K$-$\alpha$-Med2dPF.
\end{definition}
With $\alpha=1$, the $\alpha$-medoids are the medians and  $K$-$\alpha$-Med2dPF corresponds to the discrete $K$-median clustering problem in a 2d PF.
With $\alpha=2$, the $\alpha$-medoids are the canonical medoids and  $K$-$\alpha$-Med2dPF corresponds to the $K$-medoids clustering problem in a 2d PF.
Furthermore, 
Lloyd's algorithm for k-means clustering (\cite{lloyd1982least}) can be extended to solve heuristically $K$-$\alpha$-Med2dPF problems,
the case $K=2$ is the 
PAM (Partitioning Around Medoids) algorithm for k-medoids \cite{schubert2018faster}. 
We consider such neighborhood structure to define local minimums, 
the local minimums are thus characterized as following:

\begin{definition}[Local minimums of $K$-$\alpha$-Med2dPF] For all $\alpha >0$ and $K \in \NN^*$,
local minimums of $K$-$\alpha$-Med2dPF are characterized
by  the encoding of partitioning subsets $P_1,\dots,P_K$ and their respective $\alpha$-medoids $c_1,\dots,c_K$,
with the property:
\begin{equation}\label{eq::nearestCluster}
 \forall k \in [\![1;K]\!], \forall p \in P_k, \forall k' \neq k, d(p,c_k) \leqslant d(p,c_{k'})
\end{equation}

\end{definition}

\section{State-of-the-art}

This section describes   related works to appreciate our contributions, in the 
 state of the art of the k-median and k-medoid problems.

\subsection{The general p-median problem}

The p-median problem was originally a logistic problem, 
having a set of customers and defining the places of depots in order to minimize the total distance
for customers to reach the closest depot.
We give here the general form of the p-median problem.
Let $N$ be the number of clients, called $c_1 , c_2 , \dots , c_N$ , let $M$ be the number
of potential sites or facilities, called $f_1 , f_2 , \dots , f_M$ , and let $d_{i,j}$ be the distance from
$c_i$ to $f_j$. The p-median problem consists of opening $p$ facilities and assigning each
client to its closest open facility, in order to minimize the total distance.
We note that in some version of  the general p-median problem, the
graph of the possible assignments is not complete.
Considering complete grahs of distances is not a loss of generality, modeling with $d_{i,j} = + \infty$ the edges which do not exist in the original graph.

The p-median problem is naturally formulated within
the Integer Linear Programming (ILP) framework.
A first ILP formulation defines binary variables $x_{i,j} \in \{0,1\}$ and $y_{j} \in \{0,1\}$.
$x_{i,j}=1$ if and only if the customer $i$ is assigned to the  depot $j$.
$y_{j}=1$ if and only if the point $f_j$ is chosen as a depot.
Following ILP formulation expresses the p-median problem: 
\begin{equation} \label{mipPmed}
  \begin{array}{lll}
\min_{x,y} &\displaystyle\sum_{j=1}^n \sum_{i=1}^n  d_{i,j} x_{i,j} \\
  s.t: & \sum_{j=1}^n y_{j} = p \\
   & \sum_{j=1}^n x_{i,j} = 1, &    \forall i \in [\![1,n]\!] \\
& x_{i,j} \leqslant  y_{j}, &    \forall (i,j) \in [\![1,n]\!]^2,  \\
    \forall {i,j},& x_{i,j},y_{j} \in \{0,1\}  
  \end{array}
\end{equation}

The p-median problem was proven NP-hard in the general case \cite{kariv1979algorithmic}.
The p-median problem in $\RR^2$ with an Euclidian distance is also NP-hard  \cite{megiddo1984complexity}.
In a tree structure, the p-median problem is solvable in polynomial time, with a DP algorithm running in  $O(pn^2)$ time   \cite{tamir1998polynomial}.
One dimensional (1d) cases of p-median  are special cases of p-median with a tree structure, an improved
 DP algorithm was proposed with a time complexity in $O(pn)$   \cite{hassin1991improved}.

The p-median problem can be solved to optimality using ILP techniques.
The formulation (\ref{mipPmed}) is tightened in \cite{elloumi2010tighter} for a more efficient resolution  
with a Branch \& Bound (B\& B) solver.
For larger sizes of instances, Lagrangian relaxations were investigated in 
\cite{beltran2006solving,santos2009solving}
or with column-and-row generation as in \cite{avella2007computational}.
Heuristic algorithms are also widely studied \cite{mladenovic2007p}.

\subsection{From p-median to $K$-$\alpha$-Med2dPF clustering problems}

In our application, the graph is complete, the points $f_1 , f_2 , \dots , f_M$  are exactly $c_1 , c_2 , \dots , c_N$ 
and $d_{i,j}\{1/\alpha \}$  is the Euclidian distance in $\RR^2$.
 The k-medoid problem is the case $\alpha=2$, whereas the k-median problem is the case $k=1$.
Both cases are well known clustering problems.
$K$-$\alpha$-Med2dPF is not only an extension for k-median and k-medoids, the value $\alpha>0$ is a parameter of interest, varying $\alpha$ leads to different clustering solutions, as in \cite{fomin2019parameterized}.

To the best of our knowledge, no specific studies concerned  p-median or k-medoids problems in a PF before the preliminary work \cite{dupin2019medoids}.
A complexity in $O(N^3)$ time and $O(N^2)$ memory space for k-medoids problems was proven in \cite{dupin2019medoids}.
We note that an affine 2d PF is a line in $\RR^2$, such case  is equivalent to the 1d case.
 Hence, k-medoids and k-median problems are solvable in $O(kn)$ time in an affine 2d PF thanks to \cite{hassin1991improved}.

General planar cases of p-median and k-medoid problems can also be seen as specific cases of   three-dimensional (3d) PF: affine 3d PF.
Having a NP-hard complexity proven for the planar cases of p-median problems,
it implies that the corresponding p-median problems are also NP-hard for 3d PF  thanks to \cite{megiddo1984complexity}.

\subsection{Clustering/selecting points in Pareto frontiers}

We summarize here results related to the selection or the clustering of points in PF, with applications to MOO algorithms.
Maximizing the quality of discrete representations of Pareto sets was studied with  the hypervolume measure 
in the Hypervolume Subset Selection (HSS) problem \cite{auger2009investigating,sayin2000measuring}. 
 The HSS problem, maximizing the representativity of $k$ solutions among a PF of size $n$ initial ones,  is known to be NP-hard in dimension 3 (and greater dimensions) since \cite{bringmann2018maximum}.
An exact  algorithm in  $n^{O(\sqrt{k})}$ and a  polynomial-time approximation scheme for any constant dimension $d$ are also provided in \cite{bringmann2018maximum}.
The 2d case is solvable in polynomial time  thanks to  a  DP algorithm 
with a complexity in $O(kn^2)$ time and $O(kn)$ space provided in \cite{auger2009investigating}.
The time complexity of the DP algorithm  was improved in $O(kn+n \log n)$ by \cite{bringmann2014two} and in $O(k(n-k)+n \log n)$ by \cite{kuhn2016hypervolume}.

Selecting points in a 2d PF, maximizing the diversity, can be formulated also using p-dispersion problems.
Max-Min and Max-Sum p-dispersion problems are NP-hard problems\cite{erkut1990discrete,hansen1995dispersing}.
 Max-Min and Max-Sum p-dispersion problems are still NP-hard problems when distances fulfill the triangle inequality \cite{erkut1990discrete,hansen1995dispersing}.
 The planar (2d)  Max-Min p-dispersion problem is also NP-hard \cite{wang1988study}.
 The one-dimensional (1d) cases of Max-Min and Max-Sum p-dispersion problems
are solvable in polynomial time, with a similar DP algorithm running in $O(\max\{pn,n \log n\})$  time \cite{ravi1994heuristic,wang1988study}.
Max-Min p-dispersion was proven to be solvable in polynomial time, with a DP algorithm running in $O(pn \log n)$  time and $O(n)$ space 
\cite{dupin2020polynomial}.
Other variants of p-dispersion problems are also proven to be solvable in polynomial time, using also DP algorithms \cite{dupin2020polynomial}.

Some similar results exist also for k-means clustering.
K-means is NP-hard for 2d cases, and thus for 3d PF \cite{mahajan2012planar}. 
The 1d case of k-means is 
also solvable by a DP algorithm,
with a complexity in $O(kn)$ using memory  space in $O(n)$ \cite{gronlund2017fast}.
The restriction to 2d PF would be also solvable in $O(n^3)$ time with a DP algorithm if a conjecture is proven \cite{dupin2018dynamic}.

Lastly, p-center problems present also similar results.
The discrete and continuous p-center problems are NP-hard in general.
The  discrete p-center problem in $\RR^2$ with a Euclidian distance is also NP-hard \cite{megiddo1984complexity},
it implies also that the specific case of 3f PF are also NP hard using  discrete p-center culstering.
The 1d case of continuous p-center is solvable in $O(pn \log n)$ time and $O(n)$ space  \cite{megiddo1983new}, 
whereas the 2d PF cases are solvable in $O(pn \log n)$ time and $O(n)$ space 
 \cite{dupin2019planar}.
The discrete p-center problem in a 2d PF is solvable  $O(pn \log^2 n)$ time and $O(n)$ space \cite{dupin2019planar}.
 

  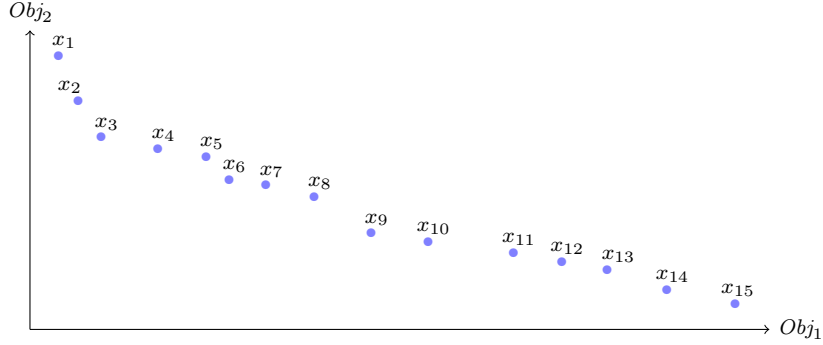
\begin{figure}[ht]
      \centering 
   \begin{tikzpicture}[scale=0.374]
    \draw[->] (0,0) -- (26,0) node[right] {$\emph{Obj}_1$};
    \draw[->] (0,0) -- (0,10.6) node[above] {$\emph{Obj}_2$};
    \draw (1.2,9.7) node[above] {$x_1$};
    \draw (1,9.7) node[color=blue!50] {$\bullet$};
    \draw (1.4,8.1) node[above] {$x_2$};
    \draw (1.7,8.1) node[color=blue!50] {$\bullet$};
    \draw (2.7,6.8) node[above] {$x_3$};
    \draw (2.5,6.8) node[color=blue!50] {$\bullet$};
    \draw (4.7,6.4) node[above] {$x_4$};
    \draw (4.5,6.4) node[color=blue!50] {$\bullet$};
    \draw (6.4,6.1) node[above] {$x_5$};
    \draw (6.2,6.1) node[color=blue!50] {$\bullet$};
    \draw (7.2,5.3) node[above] {$x_6$};
    \draw (7,5.3) node[color=blue!50] {$\bullet$}; 
    \draw (8.5,5.1) node[above] {$x_7$};
    \draw (8.3,5.1) node[color=blue!50] {$\bullet$};
    \draw (10.2,4.7) node[above] {$x_8$};
    \draw (10,4.7) node[color=blue!50] {$\bullet$};
    \draw (12.2,3.4) node[above] {$x_9$};
    \draw (12,3.4) node[color=blue!50] {$\bullet$};
    \draw (14.2,3.1) node[above] {$x_{10}$};
    \draw (14,3.1) node[color=blue!50] {$\bullet$};
    \draw (17.2,2.7) node[above] {$x_{11}$};
    \draw (17,2.7) node[color=blue!50] {$\bullet$};
    \draw (18.9,2.4) node[above] {$x_{12}$};
    \draw (18.7,2.4) node[color=blue!50] {$\bullet$};
    \draw (20.7,2.1) node[above] {$x_{13}$};
    \draw (20.3,2.1) node[color=blue!50] {$\bullet$};
    \draw (22.6,1.4) node[above] {$x_{14}$};
    \draw (22.4,1.4) node[color=blue!50] {$\bullet$};
    \draw (24.9,0.9) node[above] {$x_{15}$};
    \draw (24.8,0.9) node[color=blue!50] {$\bullet$};
    \end{tikzpicture}
    \caption{Illustration of a 2-dimensional PF with 15 points, minimizing two objectives and
     indexing the points with the Lemma \ref{reord} }\label{orderIllustr}  
   \end{figure}

\section{Intermediate results} 
   
\subsection{Local optimality and interval clustering}

In this section, local  minimums  for $K$-$\alpha$-Med2dPF clustering problems
are proven to fulfill an interval clustering property.
These results extend the previous ones, concerning only the global optimum of k-medoids problems in a 2d PF \cite{dupin2019medoids}.


\begin{lemma}
\label{transitiv}
$\preccurlyeq$ is an order relation, and 
 $\prec$ is a transitive relation:
 \begin{equation}\label{relTransitiv}
\forall x,y,z  \in \RR^2, \phantom{3} x \prec y \phantom{1} \mbox{and}\phantom{1} y \prec z \Longrightarrow x \prec z
\end{equation}
\end{lemma}

 \noindent{\textbf{Proof}: This is a  trivial consequence of  the transitivity of $\leqslant$ and $<$ in $\RR$:}

\begin{lemma}[Total order]\label{reord}
Points $(x_i)$ can be indexed such that:
\begin{eqnarray}
\forall (i_1,i_2) \in [\![1;N]\!]^2, & i_1<i_2 \Longrightarrow & x_{i_1} \prec x_{i_2}\label{ordCroissant} \\
\forall (i_1,i_2) \in [\![1;N]\!]^2, &  i_1\leqslant i_2 \Longrightarrow & x_{i_1} \preccurlyeq x_{i_2} \label{ordCroissant2}
\end{eqnarray}
This property is stronger than the property that $\preccurlyeq$ induces a total order in $E$.
Furthermore, the complexity of the sorting re-indexation is in $O(N.\log N)$
\end{lemma}

   \noindent{\textbf{Proof}}:
We reindex $E$ such that the first coordinate is increasing:
$$\forall (i_1,i_2) \in [\![1;N]\!]^2,  i_1<i_2 \Longrightarrow  x_{i_1}^1 < x_{i_2}^1$$
This sorting procedure has a complexity in $O(N.\log N)$.
Let $(i_1,i_2) \in [\![1;N]\!]^2$, with $i_1<i_2$. We have thus $x_{i_1}^1 < x_{i_2}^1$.
Having $x_{i_1} \mathcal{I} x_{i_2}$ implies $x_{i_1}^2 > x_{i_2}^2$.
$x_{i_1}^1 < x_{i_2}^1$ and $x_{i_1}^2 > x_{i_2}^2$ is by definition $x_{i_1} \prec x_{i_2}$. $\square$ 


\begin{lemma}\label{orderDist}\label{ordDist}
We suppose that points $(x_i)$ are sorted following Proposition \ref{reord}.
\begin{eqnarray}
\forall (i_1,i_2,i_3) \in [\![1;N]\!]^3,&   i_1 \leqslant i_2<i_3 \Longrightarrow d(x_{i_1},x_{i_2}) < d(x_{i_1},x_{i_3}) \label{eq1} \\
\forall (i_1,i_2,i_3) \in [\![1;N]\!]^3, &   i_1<i_2 \leqslant i_3 \Longrightarrow d(x_{i_2},x_{i_3}) < d(x_{i_1},x_{i_3}) \label{eq2}
\end{eqnarray}
\end{lemma}

\noindent{\textbf{Proof}:}
We note firstly that the equality cases are trivial, so that we can suppose $ i_1<i_2<i_3$
in the following proof. We prove here (\ref{eq1}), the proof of  (\ref{eq2}) is analogous.\\
Let $ i_1<i_2<i_3$. We note $x_{i_1}=(x^1_{i_1},x^2_{i_1})$, $x_{i_2}=(x^1_{i_2},x^2_{i_2})$  and $x_{i_3}=(x^1_{i_3},x^2_{i_3})$ .\\
Proposition \ref{reord} ordering ensures $x^1_{i_1} < x^1_{i_2} < x^1_{i_3}$ and $x^2_{i_1} > x^2_{i_2} > x^2_{i_3}$.\\
$d(x_{i_1},x_{i_2})^2 = {(x^1_{i_1} - x^1_{i_2})^2 + (x^2_{i_1} - x^2_{i_2})^2}$\\
With $x^1_{i_3} - x^1_{i_1} > x^1_{i_2} - x^1_{i_1}>0$, $(x^1_{i_1} - x^1_{i_2})^2 < (x^1_{i_1} - x^1_{i_3})^2$\\
With $x^2_{i_3} - x^2_{i_1} < x^2_{i_2} - x^2_{i_1}<0$, $(x^2_{i_1} - x^2_{i_2})^2 < (x^2_{i_1} - x^2_{i_3})^2$\\
Thus $d(x_{i_1},x_{i_2})^2 < {(x^1_{i_1} - x^1_{i_3})^2 + (x^2_{i_1} - x^2_{i_3})^2} = d(x_{i_1},x_{i_3})^2$. $\square$

\begin{prop}\label{2medOpt}[Interval clustering and local minimums of $2$-$\alpha$-Med2dPF]We suppose that points $(x_i)$ are sorted following Lemma \ref{reord}.
For all $\alpha >0$,
local minimums of $2$-$\alpha$-Med2dPF encoded with $P_1,P_2, c_1,c_2$ are necessarily on the shape 
$P_1 = \{ x_j\}_{j \in [\![1,i]\!]}$, $P_2 = \{ x_j\}_{j \in [\![i+1,N]\!]}$, far a given $i \in [\![1,N-1]\!]$.
Hence, there is at most $N-1$ local minimums for $2$-$\alpha$-Med2dPF problems.
\end{prop}
\noindent{\textbf{Proof} :} Let $P_1,P_2, c_1,c_2$ encoding a local minimum of $2$-$\alpha$-Med2dPF,
indexed such that $c_1 \prec c_2$ (thanks the total order of Lemma \ref{reord}).
Let $i = \max \{ j \in [\![1,N]\!],\: x_j \in P_1\}$.
Having $c_1 \prec c_2 \preccurlyeq x_N$, Lemma \ref{ordDist} implies that $d(x_N,c_2) \leqslant d(x_N,c_{1})$
and (\ref{eq::nearestCluster}) implies that $x_N \in P_2$ and $i<N$.
By definition of $i$, $x_{i+1} \in P_2$ and $i+1$ is the minimal index of points of $P_2$, and we have $x_{i+1}\prec  c_2$.
In terms of distances  with equation (\ref{eq::nearestCluster}),
$d(x_{i},c_1)\leqslant d(x_{i},c_2)$ and $d(x_{i+1},c_2)< d(x_{i+1},c_1)$.
We prove now that for all $i' > i+1$, $d(i',c_2)< d(i',c_1)$.
We prove by contradiction that $c_1 \preccurlyeq  x_{i}$, supposing  $x_{i} \prec c_1$. We would have thus
$x_{i+1} \preccurlyeq c_1 \prec c_2$ and  lemma \ref{ordDist} implies  $d(x_{i+1},c_1)< d(x_{i+1},c_2)$ which is in contradiction with (\ref{eq::nearestCluster}). 
Let $i' > i+1$. 
Having 
$c_1 \prec  x_{i+1}\preccurlyeq  c_2$, the total order implies $c_1 \prec x_{i+1}\preccurlyeq   c_2 \preccurlyeq  x_{i'} $
or $c_1 \prec x_{i+1}\preccurlyeq  x_{i'} \prec   c_2  $.
In the first case, $d(x_{i'},c_2)< d(x_{i'},c_1)$ is implied by lemma \ref{ordDist}.
in the second case, $d(x_{i'},c_2) \leqslant  d(x_{i+1},c_2)$ and $d(x_{i'},c_1) \geqslant  d(x_{i+1},c_1)$ using  lemma \ref{ordDist}.
Hence, using $d(x_{i+1},c_2)< d(x_{i+1},c_1)$, we have thus $d(x_{i'},c_2)\leqslant  d(x_{i+1},c_2) <  d(x_{i+1},c_1) \leqslant  d(x_{i'},c_1)$
i.e. $d(x_{i'},c_2)< d(x_{i'},c_1)$.
This proves that $P_2 = \{ x_j\}_{j \in [\![i+1,N]\!]}$, and thus $P_1 = \{ x_j\}_{j \in [\![1,i]\!]}$. $\square$

%

%
%

\begin{prop}[Optimal interval clustering]\label{propOpt}
We suppose that points $(x_i)$ are sorted following Lemma \ref{reord}.
Each local minimum of the $K$-$\alpha$-Med2dPF  problems is only composed of clusters  $\CC_{i,i'} = 
\{x_j\}_{j \in [\![i,i']\!]}= \{x \in E \: | \: \exists j \in [\![i,i']\!],\: x = x_j \} $.
As a consequence, there is at most ${N}\choose{k}$ local optima for $K$-$\alpha$-Med2dPF  problems in a 2d PF of size $N$.
\end{prop}

\noindent{\textbf{Proof}:} 
 We prove the result by contradiction on $K \in \NN$.
  We suppose having a local minimum for  $K$-$\alpha$-Med2dPF in $E$,
  encoded $P_1,\dots,P_K$ and their respective $\alpha$-medoids $c_1,\dots,c_K$,
which does not fulfill Proposition \ref{propOpt}.
There exist a cluster $k$ such that $P_k$ is not on the shape $\CC_{i,i'}$.
Denoting $i,i'$ the minimal and maximal indexes of points of $P_k$, it exists an index $j \in [\![i,i']\!]$ with $x_j \in P_{k'} \neq P_k$.
Restricting relations (\ref{eq::nearestCluster}), $P_k, P_{k'}$ must define a local minimum of $2$-$\alpha$-Med2dPF among points $E' = P_k \cup P_{k'}$.
$P_k, P_{k'}$ being nested, this is in contradiction with Proposition \ref{2medOpt}. $\square$

  \vskip 0.4cm
  
  \noindent{\textbf{Remark}: a similar property was proven for the global optima of continuous and discrete K-center clustering problems in a 2d PF \cite{dupin2019planar}, which is the key ingredient to derive a polynomial DP algorithm.
  However, a main difference exist in the optimality conditions: global optimal solution of K-center may exist with nested clusters, the optimality condition is not necessary for K-center problems. 
}

%
%



\subsection{Costs of interval clusters}

With Proposition \ref{propOpt}, the computation of cluster costs $f_{\alpha} (\CC_{i,i'})$ and the $\alpha$-medoids of interval clusters $\CC_{i,i'}$
are of special interest. We analyze here properties of these clusters to compute the costs efficiently thereafter.

 \begin{lemma}\label{inclusionLemma}
Let $P \subset P' \subset E$.
We have $f_{\alpha}(P) \leqslant f_{\alpha}(P') $.\\ 
Especially, for all $i<i'$, $f_{\alpha} (\CC_{i,i'})\leqslant f_{\alpha} (\CC_{i,i'+1})$ and $f_{\alpha} (\CC_{i+1,i'})\leqslant f_{\alpha} (\CC_{i,i'})$. 
\end{lemma}
 
 \noindent{\textbf{Proof}:} Let $P \subset P' \subset E$, and let $i$ (resp $i'$)  index defining the $\alpha$-medoids of $P$ (resp $P'$.
 Using $\sum_{x \in P}  \left|\!\left| x - x_i \right|\!\right|^{\alpha} = \min_{y \in P}  \left|\!\left| x - x_i \right|\!\right|^{\alpha}$ and the positivity of distances, we have:
\begin{eqnarray*}
 f_{\alpha}(P) & = & \sum_{x \in P}  \left|\!\left| x - x_i \right|\!\right|^{\alpha}\\
  f_{\alpha}(P) &\leqslant & \sum_{x \in P}  \left|\!\left| x - x_{i'} \right|\!\right| ^{\alpha} \\
  f_{\alpha}(P) &\leqslant & \sum_{x \in P}  \left|\!\left| x - x_{i'} \right|\!\right| + \sum_{x \in P' \setminus P}  \left|\!\left| x - x_{i'} \right|\!\right|^{\alpha} =  f_{\alpha}(P')
\end{eqnarray*}
For all $i<i'$, applying the previous result with $P=\CC_{i,i'}$ and $P=\CC_{i,i'+1}$, we have  $f_{\alpha} (\CC_{i,i'})\leqslant f_{\alpha} (\CC_{i,i'+1})$, and with $P=\CC_{i+1,i'}$ and $P=\CC_{i,i'}$ $f_{\alpha} (\CC_{i+1,i'})\leqslant f_{\alpha} (\CC_{i,i'})$. $\hfill \square$

  \vskip 0.4cm

We define  $c_{i,i'}$ as the cost of cluster $\CC_{i,i'}$ for the $K$-$\alpha$-Med2dPF clustering.
This section aims to compute efficiently the costs $c_{i,i'}$ for all $i<i'$.
By definition:
\begin{equation}
\forall i<i', \;\;\; c_{i,i'} = f_{\alpha} (\CC_{i,i'}) = \min_{j \in [\![i,i']\!]} \sum_{k \in [\![i,i']\!]}  \left|\!\left| x_j - x_k \right|\!\right|^{\alpha}
\end{equation}

We define for all $i \leqslant c \leqslant i'$, elements $d_{i,c,i'}^{\alpha}$ giving the $\alpha$-Med2dPF of cluster
$\CC_{i,i'}$ with $c$ chosen as the $\alpha$-medoid:
\begin{eqnarray}
\forall i \leqslant c \leqslant i',  \;\;\; & d_{i,c,i'}& =  \sum_{k=i}^{i'}  \left|\!\left| x_k - x_c \right|\!\right|^{\alpha} \label{defD} \\
 \forall i \leqslant i', \;\;\; & c_{i,i'}  & = \min_{l \in  [\![i,i']\!] } d_{i,l,i'} \label{defCmed}
\end{eqnarray}

   \vskip 0.4cm

  \begin{lemma}\label{interiorCentroid}
Let $i,i'$, such that $i+1<i'$. We have $d_{i,i,i'} > d_{i,i+1,i'}$ and $d_{i,i',i'} > d_{i,i'-1,i'}$.
It implies:
\begin{equation}\label{midDistCetroid}
 c_{i,i'}   = \min_{l \in  [\![i+1,i'-1]\!] } d_{i,l,i'}
\end{equation}
\end{lemma}
 
  \noindent{\textbf{Proof}:} By definition,  $d_{i,i,i'} =  \sum_{j=i}^{i'} \left|\!\left| x_j - x_i \right|\!\right|^{\alpha}$ and   
  $d_{i,i+1,i'} =  \sum_{j=i}^{i'} \left|\!\left| x_j - x_{i+1} \right|\!\right|^{\alpha}$. \\
Hence,   $d_{i,i,i'} - d_{i,i+1,i'} = \sum_{j=i+2}^{i'} \left( \left|\!\left| x_j - x_i \right|\!\right|^{\alpha}
 - \left|\!\left| x_j - x_{i+1} \right|\!\right|^{\alpha} \right)$.\\
With Lemma \ref{orderDist}, for all $j>i+1$,  $\left|\!\left| x_j - x_i \right|\!\right|^{\alpha}
 - \left|\!\left| x_j - x_{i+1} \right|\!\right|^{\alpha} >0$, which implies $d_{i,i,i'} - d_{i,i+1,i'} >0$.
 The proof of $d_{i,i',i'} > d_{i,i'-1,i'}$ is analogous, which implies also (\ref{midDistCetroid}).
 $\square$

  \vskip 0.4cm

   \begin{lemma}\label{monotonyCentroid}
Let $i,i'$, such that $i+1<i'$. Let $c\geqslant i+1 $ an index of an $\alpha$-medoid of $\CC_{i,i'}$, ie $ d_{i,c,i'} = \min_{l \in  [\![i+1,i'-1]\!] } d_{i,l,i'}$.\\
If $i'<N$, there exist $c'$ an $\alpha$-medoid of $\CC_{i,i'+1}$, such that $c \leqslant c'\leqslant i'$, i.e.  $ c_{i,i'+1}   = \min_{l \in  [\![c,i'-1]\!] } d_{i,l,i'}$.\\
If $i>1$, there exist $c''$ an $\alpha$-medoid of $\CC_{i-1,i'}$, such that $i \leqslant c'\leqslant c$, i.e. $ c_{i-1,i'}  = \min_{l \in  [\![i,c]\!] } d_{i,l,i'}$
\end{lemma}
 
  \noindent{\textbf{Proof}:}  We prove the first assertion, the last one is proven similarly. Let $i,i'$, such that $i+1<i'<N$. Let $c\geqslant i+1 $ an index of an $\alpha$-medoid of $\CC_{i,i'}$.\\
Let $c''<c$. The optimality of $c$ impose $\sum_{j=i}^{i'} \left|\!\left| x_j - x_c \right|\!\right|^{\alpha} \leqslant \sum_{j=i}^{i'} \left|\!\left| x_j - x_{c''} \right|\!\right|^{\alpha}$.
Then, Proposition  \ref{orderDist} assures that $\left|\!\left| x_{i'+1} - x_c \right|\!\right|^{\alpha} <  \left|\!\left| x_{i'+1} - x_{c''} \right|\!\right|^{\alpha}$. 
Hence, $\sum_{j=i}^{i'+1} \left|\!\left| x_j - x_c \right|\!\right|^{\alpha} < \sum_{j=i}^{i'+1} \left|\!\left| x_j - x_{c''} \right|\!\right|^{\alpha}$. In other words, any $c''<c$ is not optimal in the minimization defining  $f_{\alpha} (\CC_{i,i'+1})$ and   $c_{i,i'+1}   = \min_{l \in  [\![c,i'-1]\!] } d_{i,l,i'}$. $\hfill\square$

 \subsection{The case $1$-$\alpha$-Med2dPF }
\label{sec::singleCluster}

For all $i<i'$, the straightforward computation of $c_{i,i'}$ has a time complexity in $O((i'-i)^2)$, and thus in $O(N^2)$. 
In this section, the case of convex or concave 2d PF are proven to be solvable in  $O(N \log N)$ time when $\alpha >0$.

\begin{figure}[ht]
 \centering 
\begin{tabular}{ l }
\hline
\textbf{Algorithm 1: Computation of $ c_{i,i'}^{2}$}\\
\hline
\verb!  !  \textbf{input}: indexes $i\leqslant i'$\\
\verb!  !  \textbf{output}: the medoid cost  $c_{i,i'}^{2}$ , the index of the center
\\ 
\verb!  ! \textbf{if} $i'-i<2$ \textbf{return} $\left|\!\left| z_i - z_{i'} \right|\!\right|^2$, $i'$ \\
\verb!  ! \textbf{if} $i'-i=2$ \textbf{return} $\left|\!\left| z_{i+1} - z_{i'} \right|\!\right|^2 + \left|\!\left| z_i - z_{i+1} \right|\!\right|^2$, $i+1$ \\
\verb!  ! \\
\verb!  ! define $\mbox{idInf}=i+1$, $\mbox{valInf}=\left|\!\left| x_i - x_{i'} \right|\!\right|$, \\
\verb!  ! define $\mbox{idSup}=i'-1$, $\mbox{valSup}=\left|\!\left| x_i - x_{i'} \right|\!\right|$, \\
\verb!  ! \textbf{while} $\mbox{idSup}-\mbox{idInf}\geqslant 2$\\
\verb!    ! Compute $\mbox{idMid}= \left \lfloor \frac {i+i'} 2 \right \rfloor$, $\mbox{temp}= d_{i,\emph{idMid},i'}$, $\mbox{temp2}= d_{i,\emph{idMid}+1,i'}$\\
\verb!    ! \textbf{if} $\mbox{temp}=\mbox{temp2}$ \\
\verb!      !$\mbox{idInf}=\mbox{idMid}, \mbox{valInf}=\mbox{temp}$ \\
\verb!      !$\mbox{idSup}=1+\mbox{idMid}$, $\mbox{valSup}=\mbox{temp2}$ \\
\verb!    ! \textbf{if} $\mbox{temp}<\mbox{temp2}$ \\
\verb!      !$\mbox{idSup}=\mbox{idMid}, \mbox{valSup}=\mbox{temp}$ \\
\verb!    ! \textbf{if} $\mbox{temp}>\mbox{temp2}$ \\
\verb!      !$\mbox{idInf}=1+\mbox{idMid}, \mbox{valInf}=\mbox{temp2}$ \\
\verb!  ! \textbf{end while} \\
\textbf{if} $\mbox{valInf}< \mbox{valSup}$ \textbf{return} $\mbox{valInf}$, $\mbox{idInf}$  \\
\textbf{else return} $\mbox{valSup}$, $\mbox{idSup}$ \\

\hline
\end{tabular}
\end{figure}

\begin{prop}\label{1medoids}
We suppose that points $(z_i)$ are sorted following Lemma \ref{reord}, and that the 2d PF $E$ is convex (or concave):
it exist a $\CC^{2}$, decreasing and convex (or concave) function $g$ such that
$E = \{(x_1,g(x_1)), \dots,(x_i,g(x_i)), \dots,(x_N,g(x_N))\}$.
Then, computing $1$-$\alpha$-Med2dPF in $E$ has a  complexity in $O(N \log N)$ time and $O(1)$ additional space, using Algorithm 1. 
\end{prop}

\noindent{\textbf{Proof}:} Firstly, we notice that concave cases are implied by the convex cases, using the axial symmetry
around the line $(z_1,z_n)$. Hence, we suppose that $g$ is convex for the rest of the proof.\\
Let 
$\displaystyle h(t) = \sum_{j=1}^N (t-x_j)^{\alpha} + \sum_{j=1}^N (g(t)-g(x_j))^{\alpha}$. h is strictly convex and for all 
$j\in [\![1,N]\!]$, $h(x_j)=d_{1,j,n}$. Furthermore,  $h(x_2)<h(x_1)$ and $h(x_{n-1})<h(x_N)$ using Lemma \ref{interiorCentroid}.
Hence, $h$ is strictly decreasing, reach a minimum and then is strictly increasing.
This monotony hold  for the restrictions $h(x_j)=d_{1,j,n}$. Algorithm 1 proceeds by a dichotomic, a decreasing (resp increasing) phase detected with $d_{1,j,n}>d_{1,j+1,n}$ (resp $d_{1,j,n}<d_{1,j+1,n}$ ) implies that the optimum is greater (resp lower) than $j$. An equality $d_{1,j,n}=d_{1,j+1,n}$ implies that $j$ and $j+1$ return the optimal value. $\hfill \square$

\vskip 0.4cm
\noindent{\textbf{Remark}:} An open question is to generalize hypothesis where the Algorithm 1 is valid to compute optimally $1$-$\alpha$-Med2dPF. A counter example, with $\alpha =1$  and a PF of size $5$ is given as following:\\
$z_1 = (1,0)$, $z_2 = (0.55,0.45)$, $z_3 = (0.549,0.549)$, $z_4 = (0.45,0.45)$, $z_5 = (0,01)$.\\
$d_{1,2,5}^2=d_{1,4,5}^2= 0.45^2 + 0.55^2 +0.1^2 + 0.99^2 + 0.001^2 + 0.1^2 + 0.45^2 + 0.55^2 = 1.02002$\\
$d_{1,3,5}^2= 2 \times(0.451^2 + 0.549^2) +2 \times(0.099^2 + 0.001^2)= 1.029208$\\
We have $d_{1,1,5}>d_{1,2,5}$, $d_{1,2,5}<d_{1,3,5}$, $d_{1,3,5}>d_{1,4,5}$ and $d_{1,4,5}<d_{1,5,5}$.

\section{Computing efficiently the costs of interval clustering}

 Computing  $c_{i,i'}$ independently for all $i<i'$ using straightforward computations induces a  complexity in $O(N^4)$ time. In the case of convex (or concave) 2d PF and $\alpha >1$,
 the time complexity is in $O(N^3 \log N)$.
 \noindent{To improve the complexity, we notice that $d$ fulfills following relations:}
 \begin{equation}\label{recPmed}
  \forall i \leqslant c \leqslant i' < N, \;\;\;  d_{i,c,i'+1}  =  d_{i,c,i'} +  \left|\!\left| x_{i'+1} - x_c \right|\!\right|^{\alpha}
 \end{equation}

  \noindent{In \cite{dupin2019medoids}, relations (\ref{recPmed}) are used to compute the whole matrix of cluster costs in $O(N^4)$ time and $O(N^2)$ space. Actually, the following development will avoid to store the whole DP matrix, which induces a space complexity in  $O(N^2)$. } 
For the improved DP algorithm, we need two types of computations in $O(N)$ space:  computing the costs $f_{\alpha} (\CC_{1,j})$ for all $j \in [\![1;N]\!]$, and computing  costs $f_{\alpha} (\CC_{j',j})$  for a given $j\in [\![1;N]\!]$ and for all $j \in [\![1;j']\!]$.
  


\begin{figure}[ht]
 \centering 
\begin{tabular}{ l }
\hline
\textbf{Algorithm 2: Computing $f_{\alpha} (\CC_{1,j})$ for all $j \in [\![1;N]\!]$}\\
\hline

 \textbf{Input:} $j\in [\![1;N]\!]$, $\alpha>0$, $N$ points  of $\RR^2$, $E =\{x_1,\dots, x_N\}$\\
\textbf{Optional input:} $\overline{c}=N$ or $\overline{c}\in [\![1;N]\!]$, an upper bound to an optimal center of cluster $\CC_{1,N}$. \\
 \textbf{Output:} for all $j \in [\![1;N]\!]$, $v_{j}=f_{\alpha} (\CC_{1,j})$ and $c_j$ a center of cluster $\CC_{1,j}$\\
 
\verb!  !\\ 
\verb!  ! define vector $v$ with $v_{j}=0$  for all $j\in [\![1;N]\!]$\\ 
\verb!  ! define vector $c$ with $c_{j}=1$  for all $j\in [\![1;N]\!]$\\ 
\verb!  ! define vector $temp$ with $temp_{i}=0$  for all $i\in [\![1;N]\!]$\\ 
\verb!  ! \\
\verb!  ! \textbf{for} $j=2$ to $N$ \\ 
\verb!      !\textbf{for} $k=c_{j-1}$ to $\min(j,\overline{c})$ \\
\verb!         ! $temp_{k} =  temp_{k} +  \left|\!\left| x_k - x_{j} \right|\!\right|^{\alpha}$\\
\verb!      ! \textbf{end for} \\
\verb!      ! Compute $v_{k}  = \displaystyle\min_{k \in  [\![a,b]\!] } temp_{k}$, $c_{k}  = \mbox{arg} \displaystyle\min_{k \in  [\![a,b]\!] } temp_{k}$\\
\verb!      ! where $a=c_{j-1}$ and $b =\min(j-1,\overline{c})$\\
\verb!    ! \textbf{end for} \\
\verb!  ! \\
\textbf{return} vectors $v$ and $c$ \\
\hline
\end{tabular}
\end{figure}

\begin{prop}\label{complexClusterCostFirst}
 Algorithm 2 computes 
 the cluster costs  $f_{\alpha} (\CC_{1,j})$ for all $j \in [\![1;j]\!]$ 
with a complexity in $O(N^2)$ time and in   $O(N)$ memory space.
\end{prop}

\noindent{\textbf{Proof}:} 
To  compute the costs $f_{\alpha} (\CC_{1,j})$ for all $j \in [\![1;N]\!]$, Algorithm 2
constructs and store the costs with $j$ increasing, with a temporary vector containing the required values of $d_{1,j',j}$ at the step $j$, the step $j+1$ updating the values $d_{1,j',j+1}$ using $O(1)$ computations $d_{i,c,i'+1}  =  d_{i,c,i'} +  \left|\!\left| x_{i'+1} - x_c \right|\!\right|^{\alpha}$.
By induction, at the end of the iteration $j$, $f_{\alpha} (\CC_{1,j})$ is stored in the vector $v$ and $c$.
Even with an initial lower bound, the complexity is at least in $O(N)$ to compute each value $f_{\alpha} (\CC_{1,j})$, for a total time complexity in $O(N^2)$.
The optimal lower bounds reduces practical computations, without any improvement of the worst case complexity. The validity of the restrictions are provided by Lemma \ref{monotonyCentroid}. $\hfill\square$


\begin{figure}[ht]
 \centering 
\begin{tabular}{ l }
\hline
\textbf{Algorithm 3: Computing  $f_{\alpha} (\CC_{j',j})$ for all $j' \in [\![1;j]\!]$ for a given $j\in [\![1;N]\!]$}\\
\hline

 \textbf{Input:} $j\in [\![2;N]\!]$, $\alpha>0$, $N$ points  of $\RR^2$, $E =\{x_1,\dots, x_N\}$\\
\textbf{Optional input:} $\underline{c}=1$ or $\underline{c}\in [\![1;j]\!]$, a lower bound to an optimal center of cluster $\CC_{1,j}$. \\
 \textbf{Output:} for all $j' \in [\![1;j]\!]$, $v_{j'}=f_{\alpha} (\CC_{j',j})$ and $c_{j'}$ a center of cluster $\CC_{j',j}$\\
 
\verb!  !\\ 
\verb!  ! define vector $v$ with $v_{j'}=0$  for all $j'\in [\![1;j]\!]$\\ 
\verb!  ! define vector $c$ with $c_{j'}=j$  for all $j'\in [\![1;j]\!]$\\ 
\verb!  ! define vector $temp$ with $temp_{i}=0$  for all $i\in [\![1;j]\!]$\\ 
\verb!  ! \\
\verb!  ! \textbf{for} $j'=j-1$ to $1$ with increment $j'=j'-1$ \\ 
\verb!      !\textbf{for} $k=c_{j'+1}$ to $\max(j',\underline{c})$ with increment $k=k-1$ \\
\verb!         ! $temp_{k} =  temp_{k} +  \left|\!\left| x_k - x_{j} \right|\!\right|^{\alpha}$\\
\verb!      ! \textbf{end for} \\
\verb!      ! Compute $v_{k}  = \displaystyle\min_{k \in  [\![a,b]\!] } temp_{k}$, $c_{k}  = \mbox{arg} \displaystyle\min_{k \in  [\![a,b]\!] } temp_{k}$\\
\verb!      ! where $a=c_{j-1}$ and $b =\min(j-1,\overline{c})$\\
\verb!    ! \textbf{end for} \\
\verb!  ! \\
\textbf{return} vectors $v$ and $c$ \\
\hline
\end{tabular}
\end{figure}

\begin{prop}\label{complexCluster}\label{complexClusterCostline}
 Algorithm 3 computes 
 the cluster costs  $f_{\alpha} (\CC_{j',j})$ for all $j' \in [\![1;j]\!]$ for a given $j\in [\![1;N]\!]$ 
with a complexity in $O(j^2)$ time and in   $O(N)$ memory space.
\end{prop}

\noindent{\textbf{Proof}:} Algorithm 3 is similar with Algorithm 2:
to  compute the costs $f_{\alpha} (\CC_{j',j})$ for all $j' \in [\![1;j]\!]$ for a given $j$, Algorithm 3
constructs and store the costs with $j$ decreasing, with a temporary vector containing the required values of $d_{1,j',j}$ at the step $j$, the step $j-1$ updating the values $d_{j-1,l,j}$ using $O(1)$ computations $d_{j'-1,l,j}  =  d_{j',l,j} +  \left|\!\left| x_{j'-1} - x_c \right|\!\right|^{\alpha}$.
By induction, at the end of the iteration $j'$, $f_{\alpha} (\CC_{j',j})$ is stored in the vector $v$ and $c$.
Even with an initial upper bound, the complexity is at least in $O(j)$ to compute each value $f_{\alpha} (\CC_{j',j})$, for a total time complexity in $O(j^2)$.
The optional lower bounds reduces practical computations, without any improvement of the worst case complexity. The validity of the restrictions are provided by Lemma \ref{monotonyCentroid}. $\hfill\square$

\section{Dynamic Programming algorithm and complexity results}

in this section, the polynomial complexity of $K$-$\alpha$-Med2dPF is proven, distinguishing the case $K=2$ from the general case with $K>2$. 

\subsection{The case $2$-$\alpha$-Med2dPF}

In the case $k=2$, $2$-$\alpha$-Med2dPF can be reformulated with Proposition \ref{propOpt}, 
considering the following optimization problem instead of (\ref{defGeneric}):
\begin{equation}\label{2medDef}
 2-\alpha-Med2dPF\phantom{2}:\phantom{2} \min_{j \in [\![1;N-1]\!]} f_{\alpha} (\CC_{1,j}) + f_{\alpha} (\CC_{j+1,N})
\end{equation}
Algorithm 4 solves this reformulated problem enumerating all the possibilities following $j \in [\![1;N-1]\!]$, with an efficient computation of costs $f_{\alpha} (\CC_{1,j})$ using Algorithm 2 and $f_{\alpha} (\CC_{j+1,N})$ with Algorithm 3.
Both computations have a complexity in $O(N^2)$ time and $O(N)$ space.

\begin{figure}[ht]
 \centering 
\begin{tabular}{ l }
\hline
\textbf{Algorithm 4:  $2$-$\alpha$-Med2dPF}\\
\hline

 \textbf{Input:} $\alpha>0$, $N$ points  of $\RR^2$, $E =\{x_1,\dots, x_N\}$\\
 \textbf{Output:} $M$ the optimal value of $2$-$\alpha$-Med2dPF, and   an optimal solution\\ 
\verb!  !\\
\verb!  ! define vector $f,l$ with $f_{j}=l_{j}=0$  for all $j\in [\![1;N]\!]$\\ 
\verb!  ! initialize integers $c,j^*$ with $c,j^*=1$ and float $M=0.0$ \\ 
\verb!  ! \\
\verb!  ! run Algorithm 2: \\ 
\verb!     ! store  $f_{i} :=  f_{\alpha} (\CC_{1,i})$ for all $i \in [\![1;N-K +1]\!]$\\
\verb!     ! store $c := c_{1,N}$, $M= f_{\alpha} (\CC_{1,N-1})$\\
\verb!  ! \\
\verb!  ! run Algorithm 3 with $c$ as lower bound:\\
\verb!     !store $l_{i} :=  f_{\alpha} (\CC_{1,i})$ for all $i \in [\![1;N-K +1]\!]$\\
\verb!  ! \\
\verb!  ! \textbf{for} $j=1$ to $N-1$ \\ 
\verb!      !\textbf{if}  $f_{j} + l_{j+1} < M$  \textbf{then}  $M:=f_{j} + l_{j+1}$ and $j^*=j$\\
\verb!    ! \textbf{end for} \\
\verb!  ! \\
\textbf{return} $M$ and $[\![1;j^*]\!]$,$[\![j^* +1;N]\!]$  \\
\hline
\end{tabular}
\end{figure}
\vskip 0.4cm
\noindent{With a similar algorithm, Algorithm 4 explore all the local minima of $2$-$\alpha$-Med2dPF. Once the costs are computed in $O(N^2)$ time, the costs and $\alpha$-medoids can be stored. 
Denoting $c_i$ (resp $c'_i$) the $\alpha$-medoids of cluster $\CC_{1,i}$ (resp $\CC_{i,N}$)
$[\![1;j]\!],[\![j+1;N]\!]$ defines a local minimum if $d(c'_i,j)\geqslant d(c_i,j)$ and $d(c'_i,j+1) \leqslant d(c_i,j+1)$, such remaining computations are in $O(1)$ for each possible local optimum, and thus in $O(N)$ to test the $N-1$ possible local optimums.
Computing all the local optimums (that are at most $N-1$) has thus a complexity in $O(N^2)$.

\begin{figure}[ht]
 \centering 
\begin{tabular}{ l }
\hline
\textbf{Algorithm 4': Exhaustive search of  Local minima of  $2$-$\alpha$-Med2dPF}\\
\hline

 \textbf{Input:} $\alpha>0$, $N$ points  of $\RR^2$, $E =\{x_1,\dots, x_N\}$\\
\verb!  !\\
\verb!  ! define vector $f,l$ with $v_{j}=0$  for all $j\in [\![1;N]\!]$\\ 
\verb!  ! define vector $c,c'$ with $c_{j}=1$ and $c'_{j}=N$  for all $j\in [\![1;N]\!]$\\ 
\verb!  ! initialize integer $j^*$ with $j^*=1$ and float $M=0.0$ \\ 
\verb!  ! \\
\verb!  ! run Algorithm 2: \\ 
\verb!     ! store  $f_{i} :=  f_{\alpha} (\CC_{1,i})$ for all $i \in [\![1;N-K +1]\!]$\\
\verb!     ! store $c_{i}  := c_{1,i}$, $M= f_{\alpha} (\CC_{1,N-1})$\\
\verb!  ! \\
\verb!  ! run Algorithm 3 with $c$ as lower bound:\\
\verb!     !store $l_{i} :=  f_{\alpha} (\CC_{1,i})$ for all $i \in [\![1;N-K +1]\!]$\\
\verb!     ! store $c'_{i}  := c_{i,N}$\\
\verb!  ! \\
\verb!  ! \textbf{for} $j=1$ to $N-1$ \\ 
\verb!      !\textbf{if}  if $d(c'_i,j)\geqslant d(c_i,j)$ and $d(c'_i,j+1) \leqslant d(c_i,j+1)$\\
\verb!          !\textbf{then}  print local Minimum $[\![1;j]\!],[\![j+1;N]\!]$\\
\verb!    ! \textbf{end for} \\
\verb!  ! \\
\textbf{return}\\ 
\hline
\end{tabular}

\end{figure}

\subsection{General cases $k$-$\alpha$-Med2dPF with $k>2$}

In the general case, Proposition \ref{propOpt} allows to design a DP algorithm, enumerating the possibilities among the possibly global optimal solutions.

\begin{prop}[Bellman equations]
\label{bellmanMSN}
Defining $M_{k,i}$ as the optimal cost of  $K$-$\alpha$-Med2dPF  among the  points  indexed in $[\![1,i]\!]$ for all $k \in [\![2,K]\!]$ and  $i \in [\![k,n]\!]$,
we have:
\begin{equation}\label{initBellman}
 \forall i \in [\![1,n]\!], \:\:\: M_{1,i}=  f_{\alpha} (\CC_{1,i})
 \end{equation}
\begin{equation}\label{inducFormBellman}
\forall i \in [\![1,n]\!], \: \forall k \in [\![2,K]\!], \:\:\:  M_{k,i} = \min_{j \in [\![k-1,i-1]\!]} (M_{k-1,j} +  f_{\alpha} (\CC_{j+1,i}))
\end{equation}
\end{prop}

 \noindent{\textbf{Proof}:} (\ref{initBellman}) is trivial. We suppose $k\geqslant 2$  and prove (\ref{inducFormBellman}).
Let $ i \in [\![1,n]\!]$.
Selecting for each $j \in [\![k-1,i-1]\!]$ an optimal solution of $(k-1)$-$\alpha$-Med2dPF among points indexed in $ [\![1,j]\!]$,
and adding cluster $[\![j,i]\!]$, it makes a feasible solution for $(k-1)$-$\alpha$-Med2dPF among points indexed in $ [\![1,i]\!]$
with a cost $M_{k-1,j} +  f_{\alpha} (\CC_{j+1,i})$.
This last cost is greater than the optimal $k$-$\alpha$-Med2dPF cost, thus $M_{k,i} \leqslant  M_{k-1,j} +  f_{\alpha} (\CC_{j+1,i})$.
\begin{equation}\label{eqBellmanMSMineq}
 M_{k,i} \leqslant  \min_{j \in [\![k-1,i-1]\!]} (M_{k-1,j} +  f_{\alpha} (\CC_{j+1,i}))
\end{equation}

Let $j_1< j_2<\dots<j_{k-1}$ indexes such that $[\![1, j_1 ]\!],[\![j_1+1,j_2]\!]  ,\dots, [\![j_{k-1}+1,N]\!]$
defines an optimal solution of $k$-$\alpha$-Med2dPF, its cost is $M_{k,i}$.
Necessarily, $j_1, j_2,\dots,j_{k-2}$ defines an optimal solution of 
 $(k-1)$-$\alpha$-Med2dPF among points indexed in $ [\![1,j_{k-1}]\!]$. On the contrary, a strictly better solution for $M_{k,i}$ would be constructed adding $[\![j_{k-1}+1,N]\!]$.
We have thus:
$M_{k,i} =  M_{k-1,j_{k-1}} +   f_{\alpha} (\CC_{j_{k-1}+1,i})$.
Combined with (\ref{eqBellmanMSMineq}), it proves : $ M_{k,i} = \min_{j \in [\![k-1,i-1]\!]} (M_{k-1,j} +  f_{\alpha} (\CC_{j+1,i}))$. $\hfill \square$

\vskip 0.3cm

\begin{figure}[ht]
 \centering 
\begin{tabular}{ l }
\hline
\textbf{Algorithm 5: dynamic programming algorithm for $K$-$\alpha$-Med2dPF }\\
\hline

\textbf{Input:} \\
- $\alpha>0$ ;\\
- $N$ points  of $\RR^2$, $E =\{x_1,\dots, x_N\}$  
such that for all $ i\neq j$, $x_i \phantom{0} \mathcal{I} \phantom{0} x_j$ ;\\
- $K\in\NN^*$ the number of clusters\\
\verb!  !\\
\verb!  ! initialize  matrix $M$ with  $M_{k,i}=0$  for all $k\in [\![1;K-1]\!], i\in [\![k;N-K +k]\!]$\\
\verb!  ! initialize  vectors $u$ with  $u_{i}=\max(1,i-1)$  for all $i\in [\![1;N-1]\!]$\\
\verb!  !\\
\verb!  ! sort $E$ following the order of Proposition \ref{reord}\\
\verb!  ! run Algorithm 2: \\ 
\verb!     ! store  $M_{1,i} :=  f_{\alpha} (\CC_{1,i})$ for all $i \in [\![1;N-K +1]\!]$\\
\verb!     ! store $u_i := c_{1,i}$ for all $i \in [\![1;N]\!]$\\
\verb!  !\\
\verb!  ! \textbf{for} $i=2$ to $N-1$\\
\verb!    ! compute and store  $f_{\alpha} (\CC_{i',i})$ for all $i' \in [\![1;i]\!]$ with Algorithm 3 using $u_{i} $ as lower bound\\ 

\verb!    ! \textbf{for} $k=\max(2,K+i-N)$ to $\min(K-1,i)$ \\
\verb!      ! set $M_{k,i} = \min_{j \in [\![1,i]\!]} C_{k-1,j-1} + f_{\alpha} (\CC_{j,i})$\\
\verb!    ! \textbf{end for} \\
\verb!    ! delete the stored  $f_{\alpha} (\CC_{i',i})$ for all $i' \in [\![1;i]\!]$\\ 
\verb!  ! \textbf{end for} \\
\verb!  !\\
\verb!  ! compute and store  $f_{\alpha} (\CC_{i',N})$ for all $i' \in [\![1;N]\!]$ with Algorithm 3 using $u_{N} $ as lower bound\\
\verb!  ! set $OPT = \min_{j \in [\![2,N]\!]} M_{K-1,j-1} + f_{\alpha} (\CC_{j,N})$\\
\verb!  ! set $j = \mbox{argmin}_{j \in [\![2,N]\!]} M_{K-1,j-1} + f_{\alpha} (\CC_{j,N})$\\
\verb!  ! delete the stored  $f_{\alpha} (\CC_{i',N})$ for all $i' \in [\![1;N]\!]$\\
\verb!  !\\
\verb!  ! $i=j$\\
\verb!  ! initialize $\PP=\{[\![j;N]\!]\}$, a set of sub-intervals of $[\![1;N]\!]$.\\
\verb!  ! \textbf{for} $k=K-1$ to $2$ with increment $k \leftarrow k-1$\\
\verb!    ! compute and store  $f_{\alpha} (\CC_{i',i})$ for all $i' \in [\![1;i]\!]$\\ 
\verb!    ! find $j\in [\![1,i]\!]$ such that $M_{i,k} = M_{j-1,k-1} + f_{\alpha} (\CC_{j,i})$\\
\verb!    ! add $[\![j,i]\!]$ in $\PP$\\
\verb!    ! delete the stored  $f_{\alpha} (\CC_{i',i})$ for all $i' \in [\![1;i]\!]$\\
\verb!    ! 
$i=j-1$\\
\verb!  ! \textbf{end for} \\
\verb!  !\\
\textbf{return}  $OPT$ the optimal cost  and the partition $\PP \cup [\![1,i]\!]$\\
\hline
\end{tabular}
\end{figure}

These relations allow to compute the optimal values of $M_{k,i}$ by dynamic programming in the Algorithm 4.
$M_{K,N}$ is the optimal value of $k$-$\alpha$-Med2dPF,  backtracking  on the
matrix $(M_{k,i})_{i,k}$  computes the optimal partitioning clusters.

Actually, to compute  the optimal value $M_{K,N}$ and to recover the indexes of this optimal solution by backtracking, some elements of the matrix $(M_{k,i})_{i,k}$ are useless to compute.
For $i<N$, there is no need to compute the values $M_{K,i}$.
For $k<K$, there is no need to compute the values $M_{k,N}$.
In the line $k<K$, the elements $M_{k,i}$ for $i<k$ are not used, and $M_{k,k}=0$.
In the line $k<K$, the elements with $i>N-K+k$ will not be used in the backtracking operations starting from $M_{K,N}$.
In a line $k<K$, the indexes to compute fulfill $k<i \leqslant N-K+k$.

To compute $M_{k,i}$, it requires the previous optimal computations of  $M_{k-1,j}$ with $j<i$.
It is possible (similarly with \cite{dupin2019planar,dupin2020polynomial}), to compute the matrix $M$ following the index $k$. In Algorithm 4, the matrix $M$ is computed line by line following the index $i$ increasing.
Indeed, to compute  each value $M_{k,i}$ of the line $i$, it requires the same cluster costs $f_{\alpha} (\CC_{i',i})$ for all $i' \in [\![1;i]\!]$, computed by Algorithm 3.
It allows to use the cost computations in Algorithm 3 one time, and to delete the vector costs one the line $i$ is completed, minimizing the space memory used.
On the contrary, the first version presented in \cite{dupin2019medoids} stores the whole matrix of costs, inducing a space complexity in $O(N^2)$, whereas Algorithm 4 has a space complexity in $O(KN)$, the size of the DP matrix.

\subsection{Complexity results}

\begin{theorem}
Let  $E =\{x_1,\dots, x_N\}$ a subset of $N$ points  of $\RR^2$, 
such that for all $ i\neq j$, $x_i \phantom{0} \mathcal{I} \phantom{0} x_j$.
Clustering $E$ with $K$-$\alpha$-Med2dPF is solvable to optimality in polynomial time with Algorithm 4.
The  complexity is  $O(N^3)$ time and in $O(KN)$ memory space ,
and $O(N^2)$ time and in $O(N)$ memory space when $K=2$.
$1$-$\alpha$-Med2dPF is solvable in $O(N^2)$ time in general,
the cases of a convex or concave 2d PF with $\alpha >0$ induces a time complexity in $O(N \log N)$.

\end{theorem}

\noindent{\textbf{Proof}:}The case $K=1$ = is given in section 4.3 with Proposition \ref{1medoids}. The case $K=2$ is given using Algorithm 4 as described in section 6.1. In the following, we focus on the cases $K>2$ and Algorithm 5.
 (\ref{inducFormBellman}) uses only values $M_{k,i}$ with $j<k$ in Algorithm 5.
Induction proves that 
$M_{k,i}$ has its final value for all $i \in [\![1,N]\!]$ at the end of the for loops from $k=2$ to $K$.
$M_{k,N}$ is thus at the end of these loops the optimal value
of k-$\alpha$-Med2dPF  clustering among the $N$ points of $E$.
The backtracking phase searches for the equalities in $M_{i,k} = M_{j-1,k-1} + f_{\alpha} (\CC_{j,i})$ 
to return the optimal clusters  $\CC_{j',i}$.
Let us analyze the complexity.

Sorting and indexing the elements of $E$ following Lemma \ref{reord} has a complexity in $O(N\log N)$.
 The first line $M_{1,i}$ is computed in $O(N^2)$ time using Algorithm 2 and Proposition \ref{complexClusterCostFirst}.
 Then, to compute the line $i<N$ of the DP matrix $M_{k,i}$, the cost computations using Algorithm 3 have a complexity in $O(N^2)$ time with Proposition \ref{complexCluster},
 and the remaining operations iterating (\ref{inducFormBellman}) have a time complexity in
 $O(NK)$.
 Each line $i<N$  is thus computed in $O(N^2)$ time, the bottleneck is the  computations of cluster costs, and the total
 complexity to compute the DP matrix $M$ is  in $O(N^3)$ time. 
%
 The backtracking phase requires $K$ computations having a complexity in $O(N^2)$ time, re-computing the cluster costs, the complexity of this phase is in $O(KN^2)$ time.
 Hence, the complexity of Algorithm 5 is in $O(N^3)$ time and in $O(KN)$ memory space.  $\square$


\section{Speeding-up the DP algorithms}

In this section, it is studied how to speed-up the DP algorithms in practice, without improving the theoretical complexity proven in section 6.
On one hand, it is investigated how to remove useless computations in Algorithms 4 and 5.
On the other hand, parallelization issues are discussed.

\subsection{Additional stopping criterion}

An additional stopping criterion can be designed to stop the enumeration of cases $M_{k-1,j} +  f_{\alpha} (\CC_{j+1,i})$ to compute $M_{k,i}$:

\begin{lemma}\label{stopCriterion}
Let  $i \in [\![1,N]\!]$ and $k \in [\![2,K]\!]$. Let $\beta$ an upper bound for $M_{i,k}$.
We suppose it exist $j_0 \in [\![1,i]\!]$ such that $f_{\alpha} (\CC_{j_0,i})\geqslant  \beta$.
Then, each optimal index $j^*$ such that $M_{k,i} =   M_{k-1,j^*} +  f_{\alpha} (\CC_{j^*+1,i})$ fulfills necessarily $j^* > j_0$.
In other words, $M_{k,i} =  \min_{j \in [\![\max(k-1,j_0),i-1]\!]} M_{k-1,j} +  f_{\alpha} (\CC_{j+1,i})$
\end{lemma}
\noindent{\textbf{Proof}:} We have $f_{\alpha} (\CC_{j_0,i})\geqslant \beta \geqslant M_{k,i}$.  Lemma \ref{inclusionLemma}  implies that  for all $j<j_0$, $f_{\alpha} (\CC_{j_0,i})>f_{\alpha} (\CC_{j_0,i})\geqslant M_{k,i}$. Using $M_{k-1,j} \geqslant 0$, 
$M_{k-1,j}  + f_{\alpha} (\CC_{j_0,i})> M_{k,i}$, and thus $M_{k,i} =  \min_{j \in [\![\max(k-1,j_0),i-1]\!]} M_{k-1,j} +  f_{\alpha} (\CC_{j+1,i})$. $\hfill \square$ 
\vskip 0.4cm
Lemma \ref{stopCriterion} allows to compute optimal value of each value of the DP matrix using less computations than the algorithms proposed in the last section. A strong interest of this property is that it avoids to compute the costs $f_{\alpha} (\CC_{j,i})$ with the smallest values of $j$, ie the greatest gap $|i-j|$, which are the more time consuming cost computations. We discuss in the following how to incorporate such stopping criterion in the Algorithms 4 and 5.

\subsection{Speeding-up the case $K=2$}

In the case of $2$-$\alpha$-Med2dPF, the stopping criterion can be processed also with index $j$ increasing:

\begin{lemma}\label{stopCriterion2}
Let $\beta$ an upper bound for $M_{N,2}$.
We suppose it exist $j_1 \in [\![1,N]\!]$ such that $f_{\alpha} (\CC_{j_1,N})\geqslant  \beta$.
We suppose it exist $j_2 \in [\![1,N]\!]$ such that $f_{\alpha} (\CC_{1,j_2})\geqslant  \beta$.
Then, $M_{N,2} =  \min_{j \in [\![j_1,j_2]\!]} f_{\alpha} (\CC_{1,j}) +  f_{\alpha} (\CC_{j+1,N})$.
\end{lemma}
\noindent{\textbf{Proof}:} Lemma \ref{stopCriterion} implies that $j_1$ is a lower bound for the optimal separation index. 
Similarly,  $j_2$ is an upper bound for the optimal separation index, using that $ j \mapsto f_{\alpha} (\CC_{1,j})$ is increasing with Lemma \ref{inclusionLemma}. 
\vskip 0.4cm

\noindent{Incorporating} the stopping criterion of Lemma \ref{stopCriterion2} in Algorithm 4 shall minimize the computations of $f_{\alpha} (\CC_{1,j})$ and $f_{\alpha} (\CC_{j+1,N})$. 
Costs computations of Algorithm 2 and 3 will be proceeded only when needed.
Firstly, cost computations $f_{\alpha} (\CC_{1,j})$ (resp $f_{\alpha} (\CC_{j,N})$) are proceeded for heuristic $j \in [\![1;N/2]\!]$ (resp $j \in [\![N/2;N]\!]$), with an interruption of Algorithm 2 and 3.
Then, we set $j_1=j_2=N/2$ $\beta = f_{\alpha} (\CC_{1,N/2}) +  f_{\alpha} (\CC_{N/2+1,N})$, $\beta$ defines a feasible solution of $2$-$\alpha$-Med2dPF, and upper bound of the optimal cost.
Then, the interrupted cost computations are continued till the stopping criterion is not reached, computing costs $f_{\alpha} (\CC_{1,j}) +  f_{\alpha} (\CC_{j+1,N})$
and updating $\beta$ to the best found value which may activate earlier the stopping criterion of Lemma \ref{stopCriterion2}.

%
%
%
%
%

\subsection{Improved algorithm in the general case}

Lemma \ref{stopCriterion} can be used to stop earlier the optimal computations of each value in the DP matrix $M_{k,i}$.
For each $i \in [\![2,N]\!]$, the stopping criterion is used to stop the cost computations when all the $M_{k,i}$ for $k>2$ are proven optimal
using Lemma \ref{stopCriterion}. The advantage of such stopping criterion is to avoid the most time consuming cost computations.

Lemma \ref{stopCriterion} can also be used in a recursive  approach using memoisation. In such approach, the cost computations and some values of the DP matrix are memoised, and Lemma \ref{stopCriterion} allows to reduce the number of values to compute in the DP matrix.
However, such memoisation can induce a memory space in  $O(N^2)$.

\subsection{Parallelization issues}

A parallel implementation is a practical issue to speed up the DP algorithms.
In Algorithms 2 and 3, the inner loop induces independent computations
that can be processed in parallel, the implementation is straightforward in an environment like OpenMP.
This makes the parallelization of the crucial phase in terms of complexity in the Algorithm 4 (and also 4') for the case $K=2$. The final loop in Algorithm 2 can also be parallelized, this is less crucial than the parallelization of Algorithm 2 and 3, as this last phase  has a linear time complexity. After the improvement proposed is section 7.2, the computations in the inner loops remain independent.
For the general case, the DP algorithm as written in Algorithm 5 keeps independent computations for the inner loops to compute the costs of clusters.

\section{Conclusion and perspectives}

This paper examined properties of an extended version of the  K-medoid and K-median problems
in the special case of a discrete set of non-dominated points 
 in a two dimensional Euclidian space.
A characterization of global and local optima is proven with interval clustering.
It is proven that there is at most ${N}\choose{K}$ local minima for $K$-$\alpha$-Med2dPF  problems in a 2d PF of size $N$.
For small values of $K$, local minima can be enumerated.
The interval clustering property allows to design a  dynamic  programming algorithm with a polynomial complexity to compute the global optimal cost and a global optimal solution.
The complexity is proven  in $O(N^3)$ time and $O(KN)$ memory space when $K\geqslant 3$, cases $K=2$ having a time complexity in $O(N^2)$.
$1$-$\alpha$-Med2dPF problems are  solvable in  $O(N \log N)$ time when $\alpha>1$ the 2d PF is concave or convex.
Practical speed-up are also proposed, in relation with discussions on parallelization issues.

The complexity in $O(N^3)$ may be a bottleneck to deal with very large 2d PF, which open new perspectives.
Heuristics may apply efficiently for such cases. Initialization strategies can use optimal solutions of p-centre or p-dispersion problems in a 2d PF or in 1d cases after projection.
Having a NP-hard complexity proven for general planar cases of p-median problems, the cases of 3d PF are also NP-hard problems. 
For such cases,  perspectives are opened to design specific heuristics, but also to study approximation algorithms.

\bibliographystyle{plain}
\bibliography{biblioCluster}

\end{document}